\documentclass{bmcart}

\usepackage[utf8]{inputenc} 

\usepackage{graphicx}

\usepackage{xy}
\xyoption{matrix}
\xyoption{frame}
\xyoption{arrow}
\xyoption{arc}

\usepackage{ifpdf}
\ifpdf
\else
\PackageWarningNoLine{Qcircuit}{Qcircuit is loading in Postscript mode.  The Xy-pic options ps and dvips will be loaded.  If you wish to use other Postscript drivers for Xy-pic, you must modify the code in Qcircuit.tex}
\xyoption{ps}
\xyoption{dvips}
\fi

\entrymodifiers={!C\entrybox}

\newcommand{\qw}[1][-1]{\ar @{-} [0,#1]}

\newcommand{\gate}[1]{*+<.6em>{#1} \POS ="i","i"+UR;"i"+UL **\dir{-};"i"+DL **\dir{-};"i"+DR **\dir{-};"i"+UR **\dir{-},"i" \qw}

\newcommand{\multigate}[2]{*+<1em,.9em>{\hphantom{#2}} \POS [0,0]="i",[0,0].[#1,0]="e",!C *{#2},"e"+UR;"e"+UL **\dir{-};"e"+DL **\dir{-};"e"+DR **\dir{-};"e"+UR **\dir{-},"i" \qw}
\newcommand{\ghost}[1]{*+<1em,.9em>{\hphantom{#1}} \qw}
\newcommand{\push}[1]{*{#1}}
\newcommand{\Qcircuit}{\xymatrix @*=<0em>}

\startlocaldefs
\endlocaldefs
\usepackage{bbm,amsmath,amssymb}
\usepackage[sort&compress]{natbib}
\usepackage[matrix,frame,arrow]{xypic}

\begin{document}

\begin{frontmatter}

\begin{fmbox}
\dochead{Research}


\title{Fermionic Models with Superconducting Circuits}

\author[
   addressref={aff1},
]{\inits{ULH}\fnm{U.} \snm{Las Heras}}
\author[
   addressref={aff1},
]{\inits{LGA}\fnm{L.} \snm{Garc\'ia-\'Alvarez}}
\author[
   addressref={aff1},
]{\inits{AM}\fnm{A.} \snm{Mezzacapo}}
\author[
   addressref={aff1,aff2},
]{\inits{JRS}\fnm{E.} \snm{Solano}}
\author[
   addressref={aff1},
]{\inits{LL}\fnm{L.} \snm{Lamata}}

\address[id=aff1]{
  \orgname{Department of Physical Chemistry, University of the Basque Country UPV/EHU}, 
  \street{Apartado 644},                     %
  \postcode{48080}                                
  \city{Bilbao},                              
  \cny{Spain}                                    
}
\address[id=aff2]{%
  \orgname{IKERBASQUE, Basque Foundation for Science},
  \street{Maria Diaz de Haro 3},
  \postcode{48013}
  \city{Bilbao},
  \cny{Spain}
}



\end{fmbox}


\begin{abstractbox}

\begin{abstract} 

We propose a method for the efficient quantum simulation of fermionic systems with superconducting circuits. It consists in the suitable use of Jordan-Wigner mapping, Trotter decomposition, and multiqubit gates, be with the use of a quantum bus or direct capacitive couplings. We apply our method to the paradigmatic cases of 1D and 2D Fermi-Hubbard models, involving couplings with nearest and next-nearest neighbours. Furthermore, we propose an optimal architecture for this model and discuss the benchmarking of the simulations in realistic circuit quantum electrodynamics setups.

\end{abstract}



03.67.Ac, 42.50.Pq, 85.25.Cp, 05.30.Fk


\end{abstractbox}
%

\end{frontmatter}



\section{Background}
\label{Introduction}

Quantum simulations are one of the most promising research fields in quantum information, allowing the possibility of solving problems exponentially faster than classical computers~\cite{Feynman82}. In those cases in which analog quantum simulation is hard or impossible, one may decompose the simulated quantum dynamics in terms of discrete quantum gates through a technique known as digital quantum simulation~\cite{Lloyd96,Suzuki90,Berry07}. 
Problems involving interacting fermions are frequently intractable for classical computers due to, among other features, the exponential growth of the Hilbert space dimension with the size of the system. Moreover, standard numerical methods such as quantum Monte Carlo algorithms, do not converge for fermionic systems. Indeed, neither fermionic models in more than one dimension nor systems with the well-known sign problem~\cite{Troyer05} can be efficiently simulated employing a classical computer. Quantum simulations allow us the reproduction and study of complex systems by means of the use of minimal experimental resources and going beyond mean field approximations in numerical calculations. 

Circuit quantum electrodynamics (cQED){\cite{Devoret13} is one of the most advanced quantum technologies in terms of coherent control and scalability aspects. Several analog quantum simulations have been proposed in this quantum platform, e.g., spin models~\cite{Tsokomos10}, quantum phase transitions~\cite{Oudenaarden96}, spin glasses~\cite{Tsokomos08}, disordered systems~\cite{Ripoll08}, metamaterials~\cite{Rakhmanov08}, time symmetry breaking~\cite{Koch10}, topological order~\cite{You10}, atomic physics~\cite{Wallraff04}, open systems~\cite{Li13}, dynamical gauge theories~\cite{Marcos13}, and fermionic models in one dimension~\cite{Tian14}, among others. Furthermore, digital quantum simulations have been recently proposed for superconducting circuits~\cite{LasHeras,Mezzacapo} and two pioneering experiments have been performed~\cite{Barends, Salathe}. 

In this article, we present a method for encoding the simulation of fermionic systems for arbitrary spatial dimensions, long range or short range couplings, and highly nonlinear interactions, in superconducting circuits. For this purpose, we differentiate two kinds of cQED setups, those  employing pairwise capacitive qubit interactions~\cite{Barends14}, and the ones employing microwave resonators as quantum buses~\cite{Steffen13}. Our method can be summarized in three steps. The first step consists in mapping a set of $N$ fermionic modes to $N$ spin operators via the Jordan-Wigner transformation~\cite{Jordan28}. Then, we make use of the Trotter expansion~\cite{Lloyd96, Suzuki90, Berry07} to decompose the unitary evolution of the simulated system in a sequence of quantum gates.  Finally, many-body interactions~\cite{Molmer99} are implemented either with a sequence of capacitive two-qubit gates or by fast multiqubit gates mediated by resonators~\cite{Mezzacapo14}. Our method allows to implement highly nonlinear and long-range interactions employing only polynomial resources, which makes it suitable for simulating complex physical problems intractable for classical computers. To this extent, we analyze the simulation of the Fermi-Hubbard model with different cQED architectures, considering couplings with nearest neighbours and next-nearest neighbours in two-dimensional fermionic lattices. The structure of the article is the following. In Sec. \ref{JordanWignerTrotter}, we explain the method for decomposing a fermionic dynamics via digital techniques. In Sec. \ref{cQED Implement}, we describe the proposal for implementing the Fermi-Hubbard model in two distinct situations, either with pairwise capacitive couplings or via resonators. Finally, in Sec. \ref{Conclusions} we give our conclusions.
\section{Jordan-Wigner mapping and Trotter expansion}
\label{JordanWignerTrotter}

The Jordan-Wigner (JW) transformation allows one to map fermionic creation and annihilation operators onto spin operators. When the fermionic lattice is two or three-dimensional, it is possible that local fermionic interactions are mapped onto nonlocal spin ones. The JW mapping reads $b_k^\dagger=I_N\otimes I_{N-1}\otimes ...\otimes\sigma_{k}^+\otimes\sigma_{k-1}^z\otimes ... \otimes\sigma_{1}^z$, and $b_k=(b_k^\dagger)^\dagger$, where $b_k(b_k^{\dagger})$ are the fermionic annihilation and creation operators and $\sigma_i^\alpha$ are the spin operators of the $i$th site, being $\sigma^\alpha$ for $\alpha=x,y,z$ the Pauli matrices and $\sigma^+=(\sigma^x+i\sigma^y)/2$. 

Often, the simulating system cannot provide in a simple manner the dynamical structure of the simulated systems. Therefore, one may feel compelled to use digital methods for implementing unnatural interactions in the controllable system, based on the decomposition of the exact unitary evolution into a sequence of discrete gates~\cite{Lloyd96}. In this sense, one can use the Trotter formula~\cite{Suzuki90} in order to obtain a polynomial sequence of efficiently implementable gates. The Trotter formula is an approximation of the unitary evolution $e^{-iHt}$, where $H$ is the simulated Hamiltonian, consisting of $M$ quantum gates $e^{-iH_it}$ that fulfill the condition $H=\sum_i^M H_i$, being $H_i$ the natural interaction terms of the controllable system. The Trotter expansion can be written as $(\hbar=1)$
\begin{eqnarray}
e^{-iHt} \simeq \left(e^{-iH_1t/l} \cdots e^{-iH_Mt/l}\right)^l +\sum_{i<j}  \frac{[H_i,H_j]t^2}{2l}. \label{TrotterBasic}
\end{eqnarray} 
Here, $e^{-iH_it}$ are the gates that can be implemented in the controllable system and $l$ is the total number of Trotter steps. By shortening the execution times of the gates and applying the protocol repeatedly, the digitized unitary evolution becomes more accurate. As can be seen in Eq.~(\ref{TrotterBasic}), the error estimate in this approximation scales with $t^2/l$, such that the longer the simulated time is, the more digital steps we need to apply in order to get good fidelities.

\section{Circuit QED implementation}
\label{cQED Implement}

\subsection{Fermi-Hubbard model: small lattices with pairwise interactions}

In this section, we present a cQED encoding of the Fermi-Hubbard model, as an example of a fermionic model with nearest-neighbour pairwise interactions, which hence employs only pairwise capacitive spin-spin interactions. Although we focus on a model with three fermionic modes, for the sake of clarity, these techniques are straightforwardly extendable to arbitrary number of fermionic modes in two and three spatial dimensions. These cases are in general mapped into multi-qubit gates that can be always polynomially decomposed into sets of two-qubit gates, as shown below in Eq.~(\ref{MSDecompos}). In the last part of Sec.~\ref{cQED Implement}, we focus on another cQED platform that uses resonators instead of direct qubit couplings to mediate the interactions.

The Fermi-Hubbard dynamics is a condensed matter model describing traveling electrons in a lattice. The model captures the competition between the kinetic energy of the electrons, discretized in a lattice and encoded in a  hopping term, with their Coulomb interaction that is expressed by a nonlinear term. We begin by considering a small lattice realizable with current technology. We consider the Fermi-Hubbard-like model for three spinless fermions with open boundary conditions,
\begin{equation}
H = -h \left(b^{\dag}_1 b_2 + b^{\dag}_2 b_1 + b^{\dag}_2 b_3 + b^{\dag}_3 b_2 \right) +U \left(b^{\dag}_1 b_1 b^{\dag}_2 b_2 + b^{\dag}_2 b_2 b^{\dag}_3 b_3 \right).
\label{Hub}
\end{equation}
Here, $b_m^{\dag}$ and $b_m$ are fermionic creation and annihilation operators for the site $m$.

We will use the Jordan-Wigner transformation in our derivation to encode the fermionic operators into tensor products of Pauli matrices. We will show below that the latter may be efficiently implemented in superconducting circuits. The Jordan-Wigner mapping reads,
\begin{eqnarray}
b^{\dag}_1 &=& \mathbb{I} \otimes \mathbb{I} \otimes \sigma^{+} , \nonumber \\
b^{\dag}_2 &=& \mathbb{I} \otimes  \sigma^{+} \otimes \sigma^{z} , \nonumber \\
b^{\dag}_3 &=&  \sigma^{+} \otimes \sigma^{z} \otimes \sigma^{z}.
\end{eqnarray}
Afterwards, we rewrite the Hamiltonian in Eq.~(\ref{Hub}) in terms of spin-$1/2$ operators, 
\begin{eqnarray}
H & = & \frac{h}{2} \left(\mathbb{I} \otimes \sigma^{x} \otimes \sigma^{x} + \mathbb{I} \otimes \sigma^{y} \otimes \sigma^{y} + \sigma^{x} \otimes \sigma^{x} \otimes \mathbb{I} + \sigma^{y} \otimes \sigma^{y} \otimes \mathbb{I} \right) \nonumber \\
&& + \frac{U}{4} \big(\mathbb{I} \otimes \sigma^{z} \otimes \sigma^{z} + \mathbb{I} \otimes \sigma^{z} \otimes \mathbb{I} + \mathbb{I} \otimes \mathbb{I} \otimes \sigma^{z} + \sigma^{z} \otimes \sigma^{z} \otimes \mathbb{I} \nonumber \\
&& + \sigma^{z} \otimes \mathbb{I} \otimes \mathbb{I} + \mathbb{I} \otimes \sigma^{z} \otimes \mathbb{I}  \big).
\label{Hubspin}
\end{eqnarray}
Here, the different interactions can be simulated via digital techniques using a specific sequence of gates. We will first consider the associated Hamiltonian evolution in terms of $\exp( -i \phi \sigma^{z} \otimes \sigma^{z} )$ interactions. These can be implemented in small steps of $CZ_{\phi}$ gates, where an average single-qubit gate and two-qubit gate fidelities of 99.92\% and up to 99.4\%, respectively, have been recently achieved~\cite{Barends14}. One can then use the following relations,
\begin{eqnarray}
\sigma^{x} \otimes \sigma^{x} &=& R_{y} (\pi/2)\sigma^{z} \otimes \sigma^{z} R_{y} (-\pi/2) , \nonumber \\
\sigma^{y} \otimes \sigma^{y} &=&  R_{x} (-\pi/2)\sigma^{z} \otimes \sigma^{z} R_{x} (\pi/2) ,
\end{eqnarray}
where $R_{j} (\theta) = \exp(-i\frac{\theta}{2} \sigma^{j})$ denote local rotations along the $j$th axis of the Bloch sphere acting on both qubits.

The evolution operator associated with the Hamiltonian in Eq.~(\ref{Hubspin}) can be expressed in terms of $\exp( -i \phi \sigma^{z} \otimes \sigma^{z} )$ interactions. Moreover, the operators may be rearranged in a more suitable way in order to optimise the number of gates and eliminate global phases,
\begin{eqnarray}
\exp(-iH t) &\approx& \bigg[ R^{'}_{y} (\pi/2) \exp\left(-i \frac{h}{2} \mathbb{I} \otimes \sigma^{z} \otimes \sigma^{z} \frac{t}{n}\right) R^{'}_{y} (-\pi/2) R_{y} (\pi/2) \nonumber \\
& & \times \exp\left(-i \frac{h}{2} \sigma^{z} \otimes \sigma^{z} \otimes \mathbb{I} \frac{t}{n}\right) R_{y} (-\pi/2) R^{'}_{x} (-\pi/2) \nonumber \\
& & \times  \exp\left(-i \frac{h}{2} \mathbb{I} \otimes \sigma^{z} \otimes \sigma^{z} \frac{t}{n}\right) R^{'}_{x} (\pi/2) R_{x} (-\pi/2) \nonumber \\
& & \times \exp\left(-i \frac{h}{2} \sigma^{z} \otimes \sigma^{z} \otimes \mathbb{I} \frac{t}{n}\right) R_{x} (\pi/2) \exp\left(-i \frac{U}{4} \mathbb{I} \otimes \sigma^{z} \otimes \sigma^{z} \frac{t}{n}\right)\nonumber \\
& & \times \exp\left(-i \frac{U}{2} \mathbb{I} \otimes \sigma^{z} \otimes \mathbb{I} \frac{t}{n}\right) \exp\left(-i \frac{U}{4} \mathbb{I} \otimes \mathbb{I} \otimes \sigma^{z} \frac{t}{n}\right) \nonumber \\
& & \times \exp\left(-i \frac{U}{4} \sigma^{z} \otimes \sigma^{z} \otimes \mathbb{I} \frac{t}{n}\right) \exp\left(-i \frac{U}{4} \sigma^{z} \otimes \mathbb{I} \otimes \mathbb{I} \frac{t}{n}\right) \bigg]^{n},
\end{eqnarray}
where we use the prime notation in the rotation to distinguish between gates applied on different qubits, since $R_{i}$ acts on the first and second qubits, while $R^{'}_{i}$ acts on the second and the third. If we consider that $R_{j} (\alpha)R_{j} (\beta) = R_{j} (\alpha + \beta)$, the sequence of gates for one Trotter step in the digital simulation of the Hubbard model with three qubits is shown in Table 1.
There, gates $A$ and $B$ are two-qubit gates written in terms of $\exp( -i \phi \sigma^{z} \otimes \sigma^{z} )$ interactions, $A = \exp(-i\frac{{h}}{2}\sigma^{z} \otimes \sigma^{z}\frac{t}{n})$ and $B = \exp(-i\frac{U}{4}\sigma^{z} \otimes \sigma^{z}\frac{t}{n})$. $Z_1$ and $Z_2$ are single-qubit phases, $Z_1 = \exp(-i\frac{U}{4}\sigma^{z}\frac{t}{n})$ and $Z_2 = \exp(-i\frac{U}{2}\sigma^{z}\frac{t}{n})$, while $X_{\alpha}$ and $Y_{\alpha}$ are rotations along the $x$ and $y$ axis, respectively.

The $\exp( -i \phi \sigma^{z} \otimes \sigma^{z} )$ interaction can be implemented in small steps with optimized $CZ_{\phi}$ gates,
\[\exp\left(-i\frac{\phi}{2} \sigma^{z} \otimes \sigma^{z}\right) =
 \left( \begin{array}{cccc}
1 & 0 & 0 & 0 \\
 0 & e^{i\phi} & 0 & 0 \\
 0 & 0 & e^{i\phi} & 0 \\
 0 & 0 & 0 & 1
\end{array} \right).\]
Quantum circuits for simulating these gates are shown in Tables 2 and 3. In the Tables, $X$ and $Y$ are $\pi$ pulses.
\begin{table}[h]
\label{Table1}
\caption{Sequence of gates for one Trotter step of Hamiltonian~\ref{Hub}.} 
\centering
\[ \Qcircuit @C=.75em @R=.7em {
     & \qw & \qw &  \gate{Y_{\pi/2}} & \multigate{1}{A} & \gate{Y_{-\pi/2}} & \qw & \qw & \gate{X_{-\pi/2}} & \multigate{1}{A} & \gate{X_{\pi/2}} & \qw & \multigate{1}{B} & \gate{Z_1} & \qw \\
     & \gate{Y_{\pi/2}} & \multigate{1}{A} & \qw & \ghost{A} & \gate{Y_{-\pi/2}} & \gate{X_{-\pi/2}} & \multigate{1}{A} & \qw & \ghost{A} & \gate{X_{\pi/2}} & \multigate{1}{B} & \ghost{B} & \gate{Z_2} & \qw \\
     & \gate{Y_{\pi/2}} & \ghost{A} & \gate{Y_{-\pi/2}} & \qw & \qw & \gate{X_{-\pi/2}} & \ghost{A} & \gate{X_{\pi/2}} & \qw & \qw & \ghost{B} & \qw & \gate{Z_1} & \qw 
}\]
\end{table}
\begin{table}[h]
\label{Table2}
\caption{Two-qubit gates in terms of the optimized $CZ_{\phi}$ gate and $X$ $\pi$ pulses.} 
\centering
\[ \Qcircuit @C=1em @R=.7em {
     & \multigate{2}{e^{-i\frac{\phi}{2} \sigma^{z} \otimes \sigma^{z}}} & \qw & & & \gate{X} & \multigate{2}{CZ_{\phi}} & \gate{X} & \multigate{2}{CZ_{\phi}} & \qw & \qw \\
     & & & \push{\rule{.3em}{0em}=\rule{.3em}{0em}} & & & & & & & \\
     & \ghost{e^{-i\frac{\phi}{2} \sigma^{z} \otimes \sigma^{z}}} & \qw & & & \qw & \ghost{CZ_{\phi}} & \gate{X} & \ghost{CZ_{\phi}} & \gate{X} & \qw 
}\]
\end{table}
\begin{table}[h!]
\label{Table3}
\caption{Two-qubit gates in terms of the optimized $CZ_{\phi}$ gate and $Y$ $\pi$ pulses.} 
\centering
\[ \Qcircuit @C=1em @R=.7em {
     & \multigate{2}{e^{-i\frac{\pi + \phi}{2} \sigma^{z} \otimes \sigma^{z}}} & \qw & & & \gate{X} & \multigate{2}{CZ_{\phi}} & \gate{Y} & \multigate{2}{CZ_{\phi}} & \qw & \qw \\
     & & & \push{\rule{.3em}{0em}=\rule{.3em}{0em}} & & & & & & & \\
     & \ghost{e^{-i\frac{\pi + \phi}{2} \sigma^{z} \otimes \sigma^{z}}} & \qw & & & \qw & \ghost{CZ_{\phi}} & \gate{Y} & \ghost{CZ_{\phi}} & \gate{X} & \qw 
}\]
\end{table}
\begin{figure}[h]
\includegraphics[width=0.7 \linewidth]{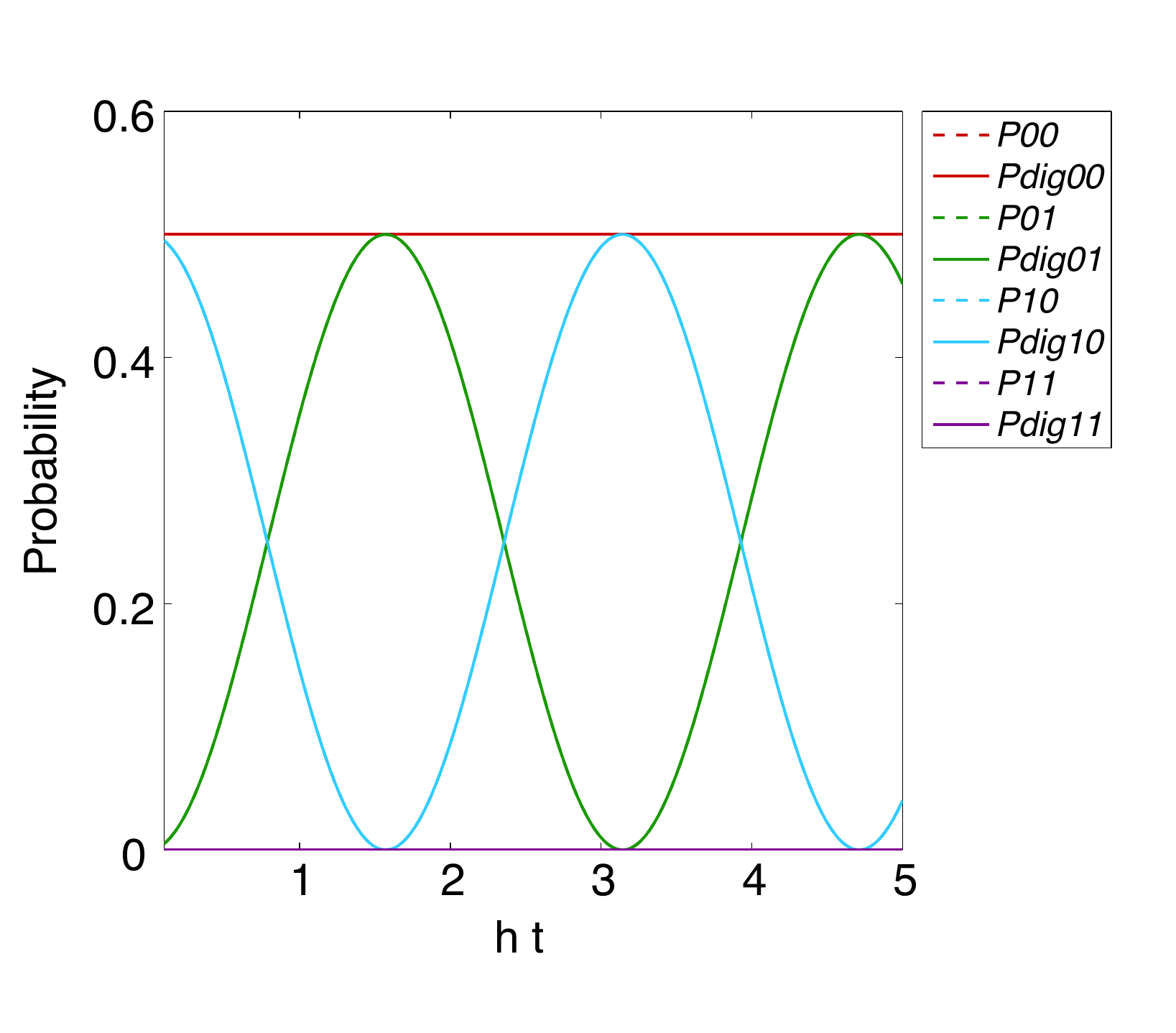} \\ 
\caption{Probability $Pij$ for each state $|ij\rangle$ in the Hubbard model with two fermionic modes. We obtain the same dynamics for hopping $h=1$, and values of the potential $U=1$ and $U=0.5$. We also consider in both cases different number of Trotter steps, $n=4$ and $n=10$, and observe the same result with no Trotter error. The initial state is in all cases $(|00\rangle + |10\rangle)/\sqrt{2}$. Dashed lines refer to numerical solutions without Trotter approximation, and solid lines to numerical solutions with Trotter approximation. The absence of Trotter error comes from the fact that the second term in the Trotter expansion in Eq.~(\ref{TrotterBasic}), i.e., the term proportional to the sum of commutators, is zero for this specific model, allowing us to perform the simulation in a single Trotter step.}
\label{Hubfig}
\end{figure}

\subsubsection{Numerical analysis of the errors}

In this section, we present numerical simulations for specific values of model parameters, that is, the time $t$, the hopping coefficient $h$, and nonlinear coupling $U$. We compute numerically the results for the proposed model with three fermionic modes, as well as the equivalent one with two fermionic modes, for the sake of completeness. In Figs.~\ref{Hubfig}~and~\ref{3Hubfig}, we show the results of the Fermi-Hubbard model with two and three fermionic modes, respectively, for $n=4$ and $n=10$ Trotter steps. As shown below, the achieved fidelities can be large at the end of each digital protocol.

In Fig.~\ref{MainFid}, we plot the fidelities of the digitally-evolved state with respect to the ideal dynamics associated with Eq.~\ref{Hub}, where $\theta\equiv Ut$, for $n=4$ Trotter steps. The fidelities are defined as  $F=|\langle\Psi_T|\Psi\rangle|^2$, being $|\Psi\rangle$ and $|\Psi_T\rangle$ the states evolved with the exact unitary evolution and the digital one, respectively. Fidelities well above 90\% can be achieved for a large fraction of the considered period.

\begin{figure}[h]
\includegraphics[width=0.98 \linewidth]{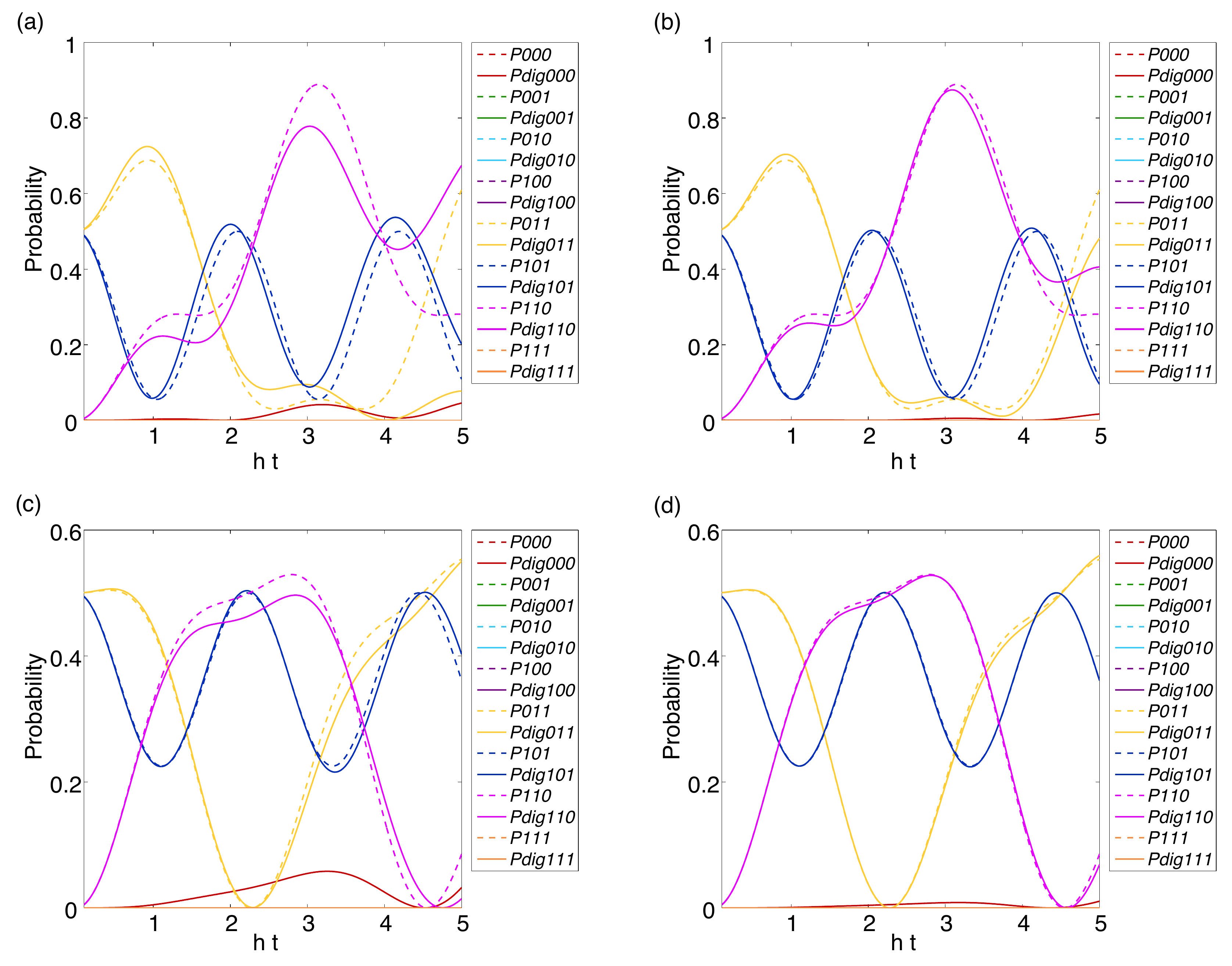} \\
\caption{Probability $Pijk$ for each state $|ijk\rangle$ in the Hubbard model with three fermionic modes. The physical parameters used are hopping $h=1$, together with (a) $U=1$ and number of Trotter steps $n=4$, (b) $U=1$ and $n=10$, (c) $U=0.5$ and $n=4$, and (d) $U=0.5$ and $n=10$. The initial state is in all cases $(|011\rangle + |101\rangle)/\sqrt{2}$. Dashed lines refer to numerical solutions without Trotter approximation, and solid lines to numerical solutions with Trotter approximation.}
\label{3Hubfig}
\end{figure}

\begin{figure}[h!]
\includegraphics[width=0.9 \linewidth]{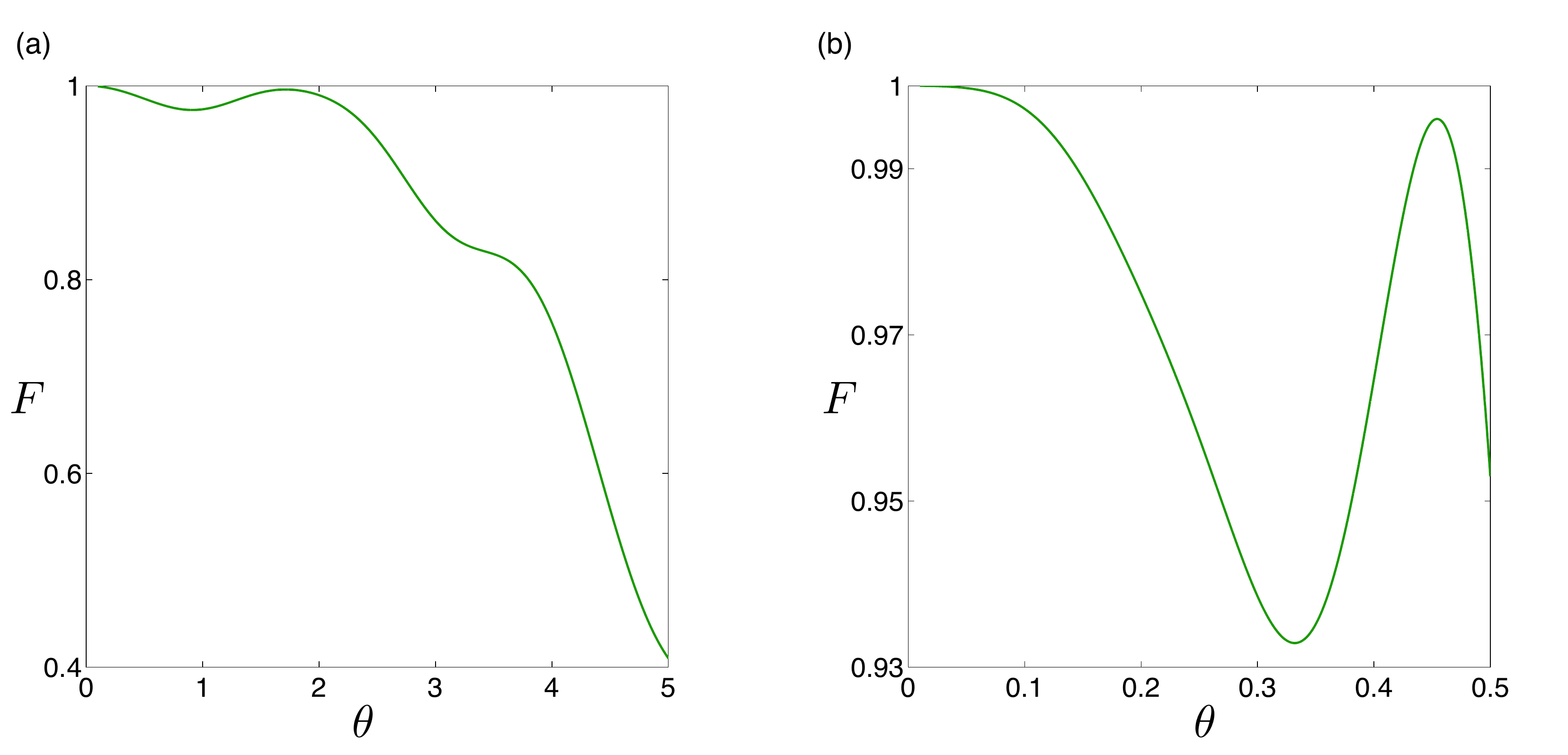}
\caption{Fidelities obtained for the dynamics of Eq.~\ref{Hub}, where $\theta\equiv U t$, for $n=4$ Trotter steps. The physical parameters used are hopping $h=1$, together with (a) $U=1$, and (b) $U=0.5$. The initial state is in both cases $(|011\rangle + |101\rangle)/\sqrt{2}$.}
\label{MainFid}
\end{figure}

Summarizing, we have analized the digital quantum simulation of the Fermi-Hubbard model with three fermionic modes in terms of simulatable spin operators with nearest neighbour interactions. Furthermore, we have considered the digital steps involving optimized gates ($CZ_\phi$).

\subsection{Large lattices and collective gates mediated via quantum bus}

Quantum simulations of fermionic and bosonic models, as well as quantum chemistry problems, have been recently proposed in trapped ions~\cite{Casanova12, Mezzacapo12, Lamata14,Yung13}. In these proposals, the use of nonlocal interactions via a quantum bus, together with digital expansion techniques, which have been implemented in recent ion-trap experimental setups~\cite{Muller11, Barreiro11}, allows for the retrieval of arbitrary fermionic dynamics.
Most current superconducting circuit setups are composed of superconducting qubits and transmission line resonators~\cite{Devoret13}. A resonator is a useful tool with several applications such as  single-qubit rotations, two-qubit gates between distant spins, and dispersive qubit readout~\cite{Blais04}. In this section, we analyze how a resonator permits the efficient reproduction of the dynamics of 2D and 3D fermionic systems.

Recently, engineering of fast multiqubit interactions with tunable transmon-resonator couplings has been proposed~\cite{Mezzacapo14}. These many-body interactions allow for the realization of multipartite entanglement~\cite{Mlyanek}, topological codes~\cite{Kitaev}, and as we show below, simulation of fermionic systems. Employing two multiqubit gates and a single-qubit rotation, the unitary evolution associated with a tensor product of spin operators can be constructed,
\begin{eqnarray}
U=U_{S_z^2}U_{\sigma_y}(\phi)U^{\dagger}_{S_z^2}=\exp[i\phi\sigma_1^y\otimes\sigma_2^z\otimes\sigma_3^z\otimes...\otimes\sigma_k^z],\label{MSDecompos}
\end{eqnarray} 
where $U_{S_z^2}=\exp[-i\pi/4\sum_{i<j}\sigma_i^z\sigma_j^z]$ and $U_{\sigma_y}(\phi)=\exp[-i\phi'\sigma_1^{y(x)}]$ for odd(even) $k$. The phase $\phi'$ also depends on the number of qubits, i.e., $\phi'=\phi$ for $k=4n+1$, $\phi'=-\phi$ for $k=4n-1$, $\phi'=-\phi$ for $k=4n-2$, and $\phi'=\phi$ for $k=4n$, where $n$ is a positive integer. Making use of this unitary evolution and introducing single qubit rotations, it is possible to generate any tensor product of Pauli matrices during a controlled phase that is given in terms of $\phi$. 

In Fig.~\ref{FigCouplings}, we show how to implement the $i$th-site hopping terms of a system made of $N$ fermionic sites onto $N$ superconducting qubits coupled to a quantum bus. Notice that local interactions between nearest and next-nearest neighbours in the square lattice involve several qubits in the experimental setup.

\begin{figure}[h!]
\includegraphics[scale=0.9]{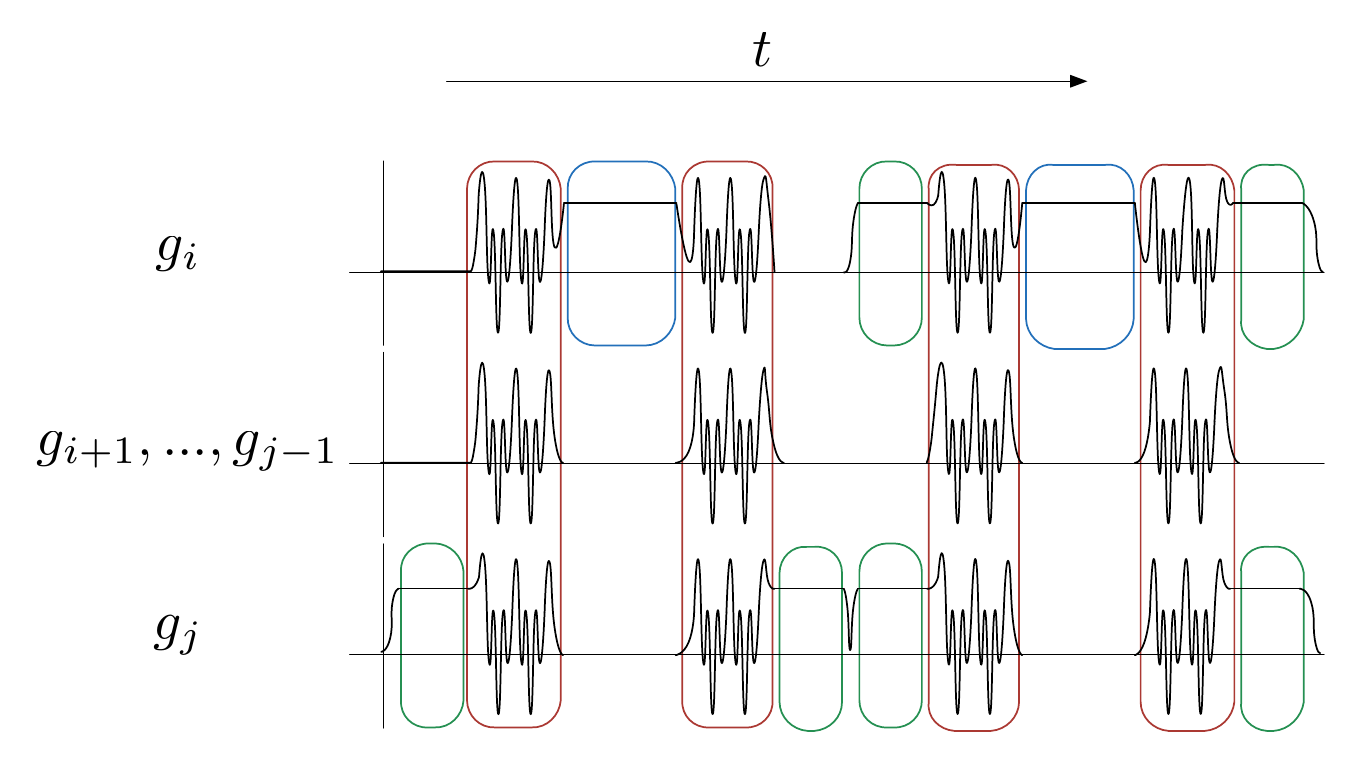}
\caption{Scheme of the magnitude of the couplings, $g_l$, of the superconducting qubits $i,...,j$ with the transmission line resonator as a function of time for the simulation of fermionic hopping terms. This sketch shows how sequences of rotations and nonlocal multiqubit gates gives place to interactions of the form $b_i^\dagger b_{j}+b_j^\dagger b_{i}$, which can be written in terms of spin operators as $-(\sigma_i^x\otimes\sigma_{i+1}^z\otimes\dots\otimes\sigma_{j-1}^z\otimes\sigma_j^x+\sigma_i^y\otimes\sigma_{i+1}^z\otimes\dots\otimes\sigma_{j-1}^z\otimes\sigma_j^y)/2$. Multiqubit gates are marked with red color where all the couplings suffer a frequency modulation~\cite{Mezzacapo14}. Single-qubit rotations are implemented by coupling a single qubit to the resonator. They are marked with green color for a phase of $\pi/4$ and with blue for a phase-dependent single-qubit rotation, $U_{\sigma_y}(\phi)$, where the phase $\phi$ is proportional to the simulated time evolution of the hopping term. Note that all the qubits between sites $i$ and $j$ play a role in this interaction in order to fulfill the Jordan-Wigner mapping. \label{FigCouplings}} 
\end{figure}

In order to benchmark our protocol with a specific example, we consider the Hamiltonian of the Fermi-Hubbard model with both nearest and next-nearest neighbour couplings,
\begin{eqnarray}
H & = & \sum_{\langle i,j \rangle}\bigg[-h(b_i^\dagger b_j+{\rm H.c.})+U\bigg(n_i-\frac{1}{2}\bigg)\bigg(n_j-\frac{1}{2}\bigg)\bigg]\nonumber\\ \label{HF}
&&+\sum_{\langle\langle i,j \rangle\rangle}\bigg[-h'(b_i^\dagger b_j+{\rm H.c.})+U'\bigg(n_i-\frac{1}{2}\bigg)\bigg(n_j-\frac{1}{2}\bigg)\bigg].
\end{eqnarray}
where $\langle  i, j \rangle$(resp., $\langle\langle i, j \rangle\rangle$) denote sum extended to nearest(next-nearest) neighbours, $h(h')$ is the hopping parameter and $U(U')$ is the interaction for nearest(next-nearest) neighbours. Here, $b_{i} (b_{i}^\dagger)$ is the fermionic annihilation(creation) operator for site $i$, that satisfies the anticommutation relation $\{b_i,b_j^\dagger\}=\delta_{i,j}$, and $n_i=b_i^\dagger b_i$ is the fermionic number operator.

Employing the method introduced before, it is possible to simulate any fermionic dynamics. Let us analyze the interactions we need to simulate in a superconducting qubit platform considering a two-dimensional lattice of $4\times4$ sites. Taking as an example the $6$th site in Fig.~\ref{FigHubbard}, the simulation of hopping terms with sites 5 and 7 requires only two-qubit gates, since they are nearest neighbours in the order chosen for the mapping. However, the simulation of hopping terms between sites 2 and 6 involves 5 superconducting qubits, $b_6^\dagger b_2$+$b_2^\dagger b_6=-1/2(\sigma_2^x\otimes\sigma_3^z\otimes\sigma_4^z\otimes\sigma_5^z\otimes\sigma_6^x+\sigma_2^y\otimes\sigma_3^z\otimes\sigma_4^z\otimes\sigma_5^z\otimes\sigma_6^y)$. The same thing happens for next-nearest neighbour interactions, which are simulated employing multiqubit gates made of either 4 or 6 spin operators. On the other hand, interaction terms between qubits $i$ and $j$ can be implemented by evolving the system with a global interaction involving all the qubits with labels between $i$ and $j$, decoupling the rest of the qubits from the resonator.

The number of gates needed for realizing this simulation depends linearly on the number of qubits. Assuming that $N$ is the number of fermionic sites in a 2D square lattice, the number of hopping and interaction terms that we need to simulate is $2\sqrt{N}(\sqrt{N}-1)$ for nearest neighbours and $2(\sqrt{N}-1)^2$ for next-nearest neighbours. As can be seen in Fig.~\ref{FigCouplings}, every  hopping term involving qubits with distant labels is made of 8 single-qubit rotations and 4 multiqubit gates. On the other hand, interaction terms can be simulated applying just one multiqubit gate.

The superconducting setup that we are considering for this quantum simulation is composed of $N$ transmon qubits coupled to a single resonator. In order to perform highly nonlocal interactions between two distant qubits, every qubit with label inside the interval spanned by them should interact with the same resonator. Coupling several qubits to just one resonator can be a difficult task wherever the number of simulated sites is large enough. Therefore, we propose an optimized architecture for the simulation of Fermi-Hubbard model with up to next-nearest neighbours in~2D. As it is shown in Fig.~\ref{FigHubbard}, we propose a setup with $N$ superconducting circuits distributed in a square lattice~\cite{Helmer}. Sequentially coupling two rows by a single transmission line resonator, one can reduce the number of qubits coupled to a single resonator. Nevertheless, all the interactions needed for satisfying the Jordan-Wigner mapping can be simulated with this architecture. Furthermore, one can achieve a speedup of the protocol by performing interactions that involve different qubits in a parallel way, e.g. the interaction between qubits 2 and 3 and the one between qubits 5 and 9 can be performed simultaneously using resonators 1 and 2, respectively.

\begin{figure}[h!]
\includegraphics[scale=0.35]{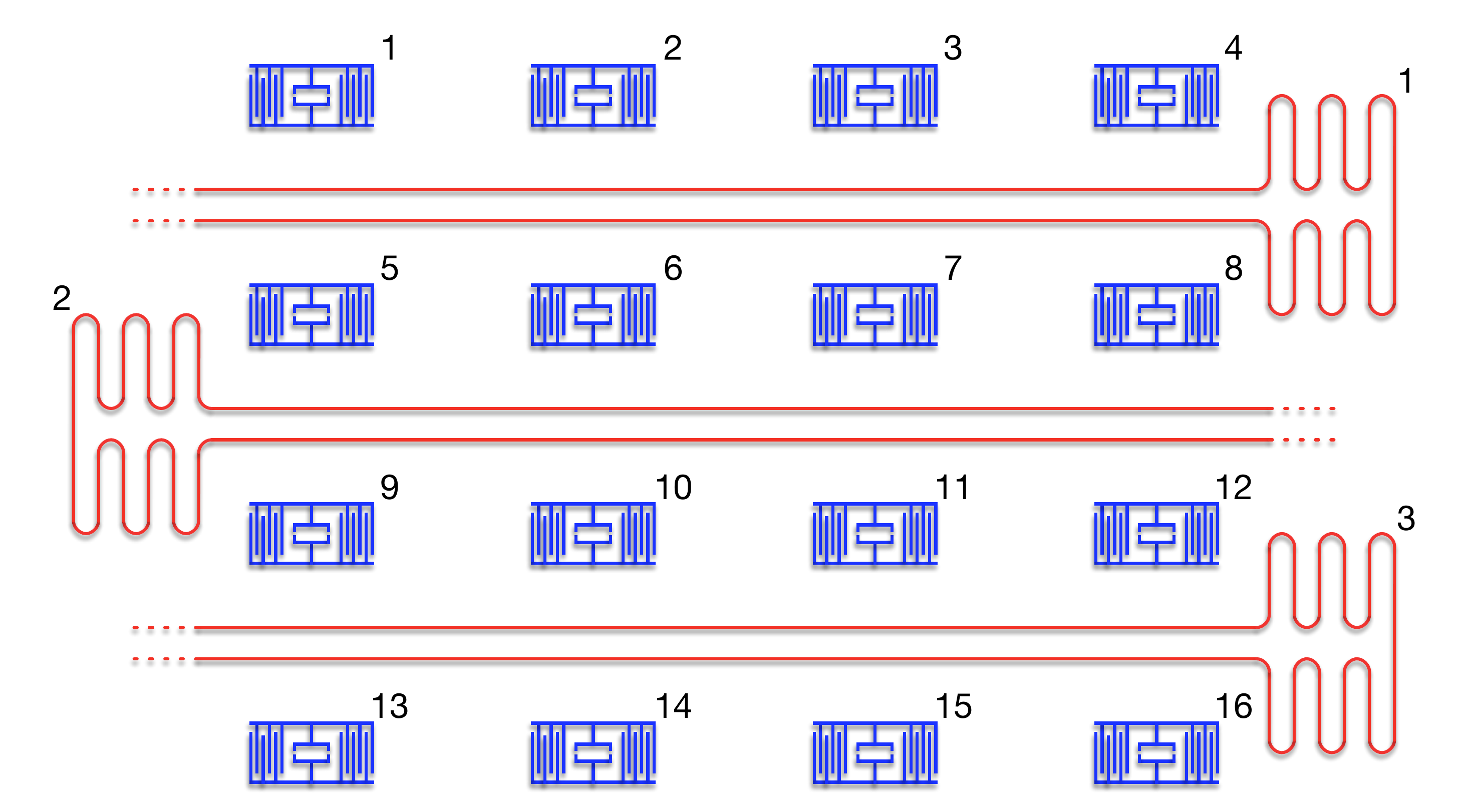}
\caption{Architecture of superconducting qubits coupled to microwave resonators  optimized for the simulation of a square lattice of $4\times4$ fermionic sites with Fermi-Hubbard interactions between nearest and next-nearest neighbours. By the use of resonators as quantum buses in the dispersive regime, several qubits are coupled allowing the implementation of single and many-body gates, which are necessary for the simulation of fermionic operators. Coupling two subsequent rows of superconducting qubits via a resonator allows to implement all the interactions required for the simulation. In order to scale the system, one needs to couple two more qubits to every resonator for simulating one more column of sites, or make use of one more resonator for connecting another row. This scheme shows an optimized architecture for the simulation of fermionic models, and further resonators would be required for the read-out and single-control of the qubits.
\label{FigHubbard}} 
\end{figure}
\begin{figure}[h!]
\includegraphics[scale=0.65]{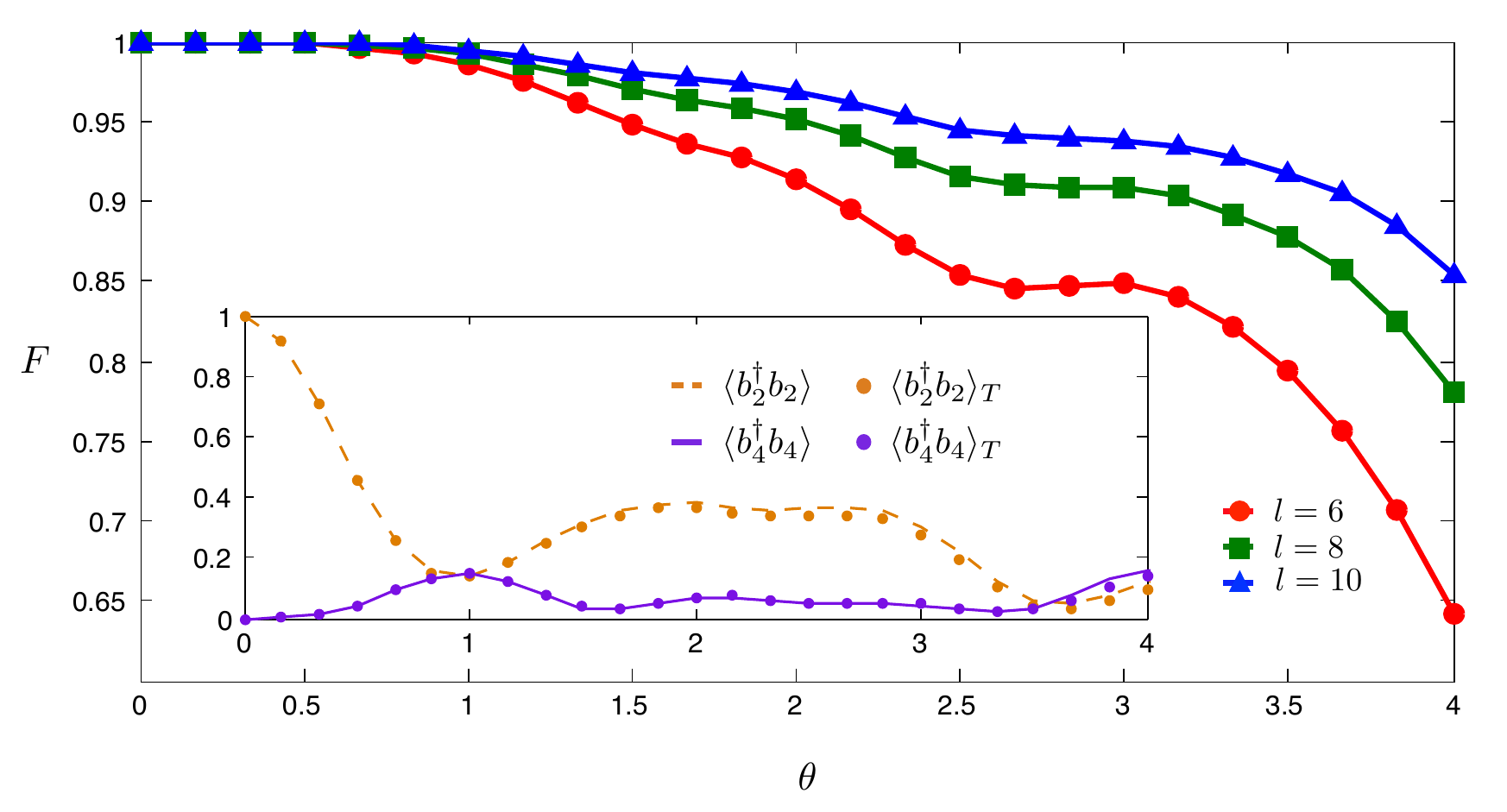}
\caption{Simulation of a 2D lattice of $3\times3$ sites where the coupling ratios are: $U/t=2$, $U/t'=10$ and $U/U'=5$. The principal plot shows the fidelity of the evolved state with digital methods for a phase of $\theta\equiv Ut=4$ applying different numbers of Trotter steps, $l$. The inset shows the population of sites 2 and 4 being the initial state $|\Psi_I\rangle=b^\dagger_2|0\rangle$. The markers denote the digital evolution with 10 Trotter steps while the lines show the exact evolution.
\label{FigFidelityPopulation}}
\end{figure}

In order to benchmark our protocol, we study its efficiency by computing the error associated with a digital decomposition. To this extent, we analyze the occupation of the fermionic sites in a $3\times3$ lattice. In Fig.~\ref{FigFidelityPopulation}, we show a plot of these populations considering a perfect unitary evolution of the Fermi-Hubbard model versus the evolution associated with the digital decomposition, where $l$ is the number of Trotter steps. As $l$ increases, the fidelity $F=|\langle\Psi_T|\Psi\rangle|^2$ improves, being $|\Psi\rangle$ and $|\Psi_T\rangle$ the states evolved with the exact unitary evolution and the digital one, respectively.

\section{Conclusions}
\label{Conclusions}

We have presented a method for the digital quantum simulation of many-body fermionic systems in superconducting circuits with polynomial resources. Moreover, we have analyzed the efficiency of this method for the simulation of the Fermi-Hubbard model in 1D and 2D with different superconducting platforms. Finally, we have proposed an optimized circuit QED architecture where our ideas may be implemented. This work paves the way for the quantum simulation of complex fermionic dynamics in superconducting circuits.

\begin{backmatter}

\section*{Competing interests}
  The authors declare that they have no competing interests.

\section*{Author's contributions}
    All the authors contributed to the realization and writing of the manuscript. 

\section*{Acknowledgments}
\label{Acknowledgments}
The authors acknowledge useful discussions with John Martinis group at Google/UCSB. We also acknowledge funding from Basque Government IT472-10 Grant, Spanish MINECO FIS2012-36673-C03-02, Ram\'on y Cajal Grant RYC-2012-11391, UPV/EHU PhD grant, UPV/EHU Project No. EHUA14/04, UPV/EHU UFI 11/55, CCQED, PROMISCE and SCALEQIT European projects.

\end{backmatter}


\begin{thebibliography}{}

\bibitem{Feynman82}Feynman RP: {\bf Simulating physics with computers}. {\it Int. J. Theor. Phys.} 1982, {\bf 21}:467.

\bibitem{Lloyd96} Lloyd S: {\bf Universal quantum simulators}. {\it Science} 1996, {\bf 273}:1073.

\bibitem{Suzuki90} Suzuki M: {\bf Fractal decomposition of exponential operators with applications to many-body theories and Monte Carlo simulations}. {\it Phys. Lett. A} 1990, {\bf 146}:319.

\bibitem{Berry07} Berry DW, Ahokas G, Cleve R, Sanders BC: {\bf Efficient quantum algorithms for simulating sparse Hamiltonians}. {\it Commun. Math. Phys.} 2007, {\bf 270}:359.

\bibitem{Troyer05} Troyer M, Wiese U-J: {\bf Computational complexity and fundamental limitations to Fermionic quantum monte carlo simulations}. {\it Phys. Rev. Lett.} 2005, {\bf 94}:170201.

\bibitem{Devoret13} Devoret MH, Schoelkopf RJ: {\bf Superconducting circuits for quantum information: an outlook}. {\it Science} 2013, {\bf 339}:1169.

\bibitem{Tsokomos10} Tsomokos DI, Ashhab S, Nori F: {\bf Using superconducting qubit circuits to engineer exotic lattice systems}. {\it Phys. Rev. A.} 2010, {\bf 82}:052311.

\bibitem{Oudenaarden96} van Oudenaarden A, Mooij JE: {\bf One-dimensional Mott insulator formed by quantum vortices in Josephson junction arrays}. {\it Phys. Rev. Lett.} 1996, {\bf 76}:4947.

\bibitem{Tsokomos08} Tsomokos DI, Ashhab S, Nori F: {\bf Fully connected network of superconducting qubits in a cavity}. {\it New J. Phys.} 2008, {\bf 10}:113020.

\bibitem{Ripoll08} Garc\'ia-Ripoll JJ, Solano E, Martin-Delgado MA: {\bf Quantum simulation of Anderson and Kondo lattices with superconducting qubits}. {\it Phys. Rev. B} 2008, {\bf 77}:024522.

\bibitem{Rakhmanov08} Rakhmanov AL, Zagoskin AM, Savel'ev S, Nori F: {\bf Quantum metamaterials: Electromagnetic waves in a Josephson qubit line}. {\it Phys. Rev. B} {\bf 77}:144507.

\bibitem{Koch10} Koch J, Houck AA, Hur KL, Girvin SM: {\bf Time-reversal-symmetry breaking in circuit-QED-based photon lattices}. {\it Phys. Rev. A.} 2010, {\bf 82}:043811.

\bibitem{You10} You JQ, Shi X-F, Hu X, Nori F: {\bf Quantum emulation of a spin system with topologically protected ground states using superconducting quantum circuits}. {\it Phys. Rev. B} 2010, {\bf 81}:014505.

\bibitem{Wallraff04} Wallraff A, Schuster DI, Blais A, Frunzio L, Huang R-S, Majer J, Kumar S, Girvin SM, Schoelkopf RJ: {\bf Strong coupling of a single photon to a superconducting qubit using circuit quantum electrodynamics}. Nature 2004, {\bf 431}:162.

\bibitem{Li13} Li J, Silveri MP, Kumar KS, Pirkkalainen J-M, Veps\"al\"ainen A, Chien WC, Tuorila J, Sillanp\"a\"a MA, Hakonen PJ, Thuneberg EV, Paraoanu GS: {\bf Motional averaging in a superconducting qubit}. {\it Nat. Commun.} 2013, {\bf 4}:1420.

\bibitem{Marcos13} Marcos D, Rabl P, Rico E, Zoller P: {\bf Superconducting circuits for quantum simulation of dynamical gauge fields}. {\it Phys. Rev. Lett.} 2013, {\bf 111}:110504.

\bibitem{Tian14} Stojanovic VM, Vanevic M, Demler E, Tian L: {\bf Transmon-based simulator of nonlocal electron-phonon coupling: A platform for observing sharp small-polaron transitions}. {\it Phys. Rev. B} 2014, {\bf 89}:144508.

\bibitem{LasHeras} Las Heras U, Mezzacapo A, Lamata L, Filipp S, Wallraff A, Solano E: {\bf Digital quantum simulation of spin systems in superconducting circuits}. {\it Phys. Rev. Lett.} 2014, {\bf 112}:200501.

\bibitem{Mezzacapo} Mezzacapo A, Las Heras U, Pedernales JS, DiCarlo L, Solano E, Lamata L: {\bf Digital quantum Rabi and Dicke models in superconducting circuits}. {\it Sci. Rep.} 2014, {\bf 4}:7482.

\bibitem{Barends} Barends R, Lamata L, Kelly J, Garc\'ia-\'Alvarez L, Fowler AG, Megrant A, Jeffrey E, White TC, Sank D, Mutus JY, Campbell B, Chen Y, Chen Z, Chiaro B, Dunsworth A, Hoi I-C, Neill C, O'Malley PJJ, Quintana C, Roushan P, Vainsencher A, Wenner J, Solano E, Martinis JM: {\bf Digital quantum simulation of fermionic models with a superconducting circuit}. arXiv:1501.07703.

\bibitem{Salathe} Salath\'e Y, Mondal M, Oppliger M, Heinsoo J, Kurpiers P, Poto\u cnik A, Mezzacapo A, Las Heras U, Lamata L, Solano E, Filipp S, Wallraff A: {\bf Digital quantum simulation of spin models with circuit quantum electrodynamics}. arXiv:1502.06778.

\bibitem{Barends14} Barends R, Kelly J, Megrant A, Veitia A, Sank D, Jeffrey E, White TC, Mutus J, Fowler AG, Campbell B, Chen Y, Chen Z, Chiaro B, Dunsworth A, Neill C, O'Malley P, Roushan P, Vainsencher A, Wenner J, Korotkov AN, Cleland AN, Martinis JM: {\bf Superconducting quantum circuits at the surface code threshold for fault tolerance}. {\it Nature} 2014, {\bf 508}:500.

\bibitem{Steffen13} Steffen L, Salathe Y, Oppliger M, Kurpiers P, Baur M, Lang C, Eichler C, Puebla-Hellmann G, Fedorov A, Wallraff A: {\bf Deterministic quantum teleportation with feed-forward in a solid state system}. {\it Nature} 2013, {\bf 500}:319.

\bibitem{Jordan28} Jordan P, Wigner E: {\bf \"Uber das Paulische \"Aquivalenzverbot}. {\it Z. Phys.} 1928, {\bf 47}:631.

\bibitem{Molmer99}M\o lmer K, S\o rensen A: {\bf Multiparticle entanglement of hot trapped ions}. {\it Phys. Rev. Lett.} 1999, {\bf 82}:1835; 

\bibitem{Sorensen00}S\o rensen A, M\o lmer K: {\bf Entanglement and quantum computation with ions in thermal motion}. {\it Phys. Rev. A.} 2000, {\bf 62}:022311.

\bibitem{Mezzacapo14} Mezzacapo A, Lamata L, Filipp S, Solano E: {\bf Many-body interactions with tunable-coupling transmon qubits}. {\it Phys. Rev. Lett.} 2014, {\bf 113}:050501.

\bibitem{Casanova12} Casanova J, Mezzacapo A, Lamata L, Solano E: {\bf Quantum simulation of interacting fermion lattice models in trapped ions}. {\it Phys. Rev. Lett.} 2012, {\bf 108}:190502.

\bibitem{Mezzacapo12} Mezzacapo A, Casanova J, Lamata L, Solano E: {\bf Digital quantum simulation of the Holstein model in trapped ions}. {\it Phys. Rev. Lett.} 2012, {\bf 109}:200501.

\bibitem{Lamata14} Lamata L, Mezzacapo A, Casanova J, Solano E: {\bf Efficient quantum simulation of fermionic and bosonic models in trapped ions}. {\it EPJ Quantum Technology} 2014, {\bf 1}:9.

\bibitem{Yung13} Yung M-H, Casanova J, Mezzacapo A, McClean J, Lamata L, Aspuru-Guzik A, Solano E: {\bf From transistor to trapped-ion computers for quantum chemistry}. {\it Sci. Rep.} 2014, {\bf 4}:3589.

\bibitem{Muller11} M\"uller M, Hammerer K, Zhou YL, Roos CF, Zoller P: {\bf Simulating open quantum systems: from many-body interactions to stabilizer pumping}. {\it New J. Phys.} 2011, {\bf 13}:085007.

\bibitem{Barreiro11} Barreiro JT, M\"uller M, Schindler P, Nigg D, Monz T, Chwalla M, Hennrich M, Roos CF, Zoller P, Blatt R: {\bf An open-system quantum simulator with trapped ions}. {\it Nature} 2011, {\bf 470}:486.

\bibitem{Blais04} Blais A, Huang RS, Wallraff A, Girvin SM, Schoelkopf RJ: {\bf Cavity quantum electrodynamics for superconducting electrical circuits: An architecture for quantum computation}. {\it Phys. Rev. A.} 2004, {\bf 69}:062320; 

\bibitem{Blais07} Blais A, Gambetta J, Wallraff A, Schuster DI, Girvin SM, Devoret MH, Schoelkopf RJ: {\bf Quantum-information processing with circuit quantum electrodynamics}. {\it Phys. Rev. A.} 2007, {\bf 75}:032329.

\bibitem{Mlyanek} Mlynek JA, Abdumalikov AA Jr, Fink JM, Steffen L, Baur M, Lang C, van Loo AF, Wallraff A: {\bf Demonstrating W-type entanglement of Dicke states in resonant cavity quantum electrodynamics}. {\it Phys. Rev. A.} 2012, {\bf 86}:053838.

\bibitem{Kitaev} Kitaev AY: {\bf Fault-tolerant quantum computation by anyons}. {\it Ann. Phys.} 2003, {\bf 303}:2.

\bibitem{Helmer} Helmer F, Mariantoni M, Fowler AG, von Delft J, Solano E, Marquardt F: {\bf Cavity grid for scalable quantum computation with superconducting circuits}. {\it Europhys. Lett.} 2009, {\bf 85}:50007.


\end{thebibliography}
\end{document}